\documentclass[conference]{IEEEtran}

\usepackage{amsmath,amssymb,amsfonts}
\usepackage{algorithmic}
\usepackage{graphicx}
\usepackage{textcomp}
\usepackage{xcolor}
\def\BibTeX{{\rm B\kern-.05em{\sc i\kern-.025em b}\kern-.08em
    T\kern-.1667em\lower.7ex\hbox{E}\kern-.125emX}}
    
\PassOptionsToPackage{table,xcdraw}{xcolor}

\AtBeginDocument{%
  \providecommand\BibTeX{{%
    Bib\TeX}}}

\usepackage[english]{babel}
\usepackage[utf8]{inputenc}

\usepackage{flushend}
\usepackage{balance}
\usepackage{multirow}

\usepackage{color}
\usepackage{booktabs}
\usepackage{fancybox}
\usepackage{xcolor}
\usepackage{float}
\usepackage{array,graphicx}
\usepackage[flushleft]{threeparttable}
\usepackage[most]{tcolorbox}
\usepackage{xspace}
\usepackage{listings}
\usepackage{pifont}
\usepackage{tikz}
\usetikzlibrary{shapes.geometric}
\usepackage{blindtext}
\usepackage{colortbl}
\usepackage[table,xcdraw]{xcolor}

\usepackage{makecell}
\usepackage{stfloats}
\usepackage{url}

\usepackage{comment}

\definecolor{purplish}{HTML}{D8D0E3}
\definecolor{purplishlight}{HTML}{EBE7F1}
\definecolor{purplishdark}{HTML}{20aee5}

\newtcolorbox[auto counter,number within=section]{rqbox}[2]{
    nameref=#1,
    title=\small{#1}, 
    enhanced,
    attach boxed title to top left={yshift=-6pt, xshift=8pt},
    boxed title style={size=small,boxsep=1pt},
    colframe=purplishdark,colback=white,colbacktitle=purplishdark,
    boxsep=2pt,left=2pt,right=2pt,top=6pt,bottom=2pt,middle=2pt
}

\newcommand{\rqtextone}{What inclusivity bugs does GitHub pose for newcomers trying to make their first contribution?}
\newcommand{\rqone}[2][]{
    \begin{rqbox}{\textbf{Research Question 1}}{#2}
        \rqtextone
        #1
    \end{rqbox}
}
\newcommand{\rqtexttwo}{What are the effects of fixing those inclusivity bugs?}
\newcommand{\rqtwo}[2][]{
    \begin{rqbox}{\textbf{Research Question 2}}{#2}
        \rqtexttwo
        #1
    \end{rqbox}
}
    
\begin{document}

\title{Designing for Cognitive Diversity: Improving the GitHub Experience for Newcomers}

%\author{\IEEEauthorblockN{Anonymous}
%\IEEEauthorblockA{Anonymous}
%Email: anonymous}

\author{\IEEEauthorblockN{Italo Santos\IEEEauthorrefmark{1}, João Felipe Pimentel\IEEEauthorrefmark{1}, Igor Wiese\IEEEauthorrefmark{2}, Igor Steinmacher\IEEEauthorrefmark{1}, Anita Sarma\IEEEauthorrefmark{3} and Marco A. Gerosa\IEEEauthorrefmark{1}}
\IEEEauthorblockA{\IEEEauthorrefmark{1}Northern Arizona University, Flagstaff, AZ, USA\\}
\IEEEauthorblockA{\IEEEauthorrefmark{2}Federal University of Technology, Campo Mourao, PR, Brazil\\}
\IEEEauthorblockA{\IEEEauthorrefmark{3}Oregon State University, Corvallis, OR, USA\\}
Email: italo\_santos@nau.edu, joao.pimentel@nau.edu, igor@utfpr.edu.br, \\igor.steinmacher@nau.edu, anita.sarma@oregonstate.edu, marco.gerosa@nau.edu}

\maketitle

\begin{abstract}
    Social coding platforms such as GitHub have become \textit{defacto} environments for collaborative programming and open source. When these  platforms do not support specific cognitive styles, they create barriers to programming for some populations. Research shows that the cognitive styles typically favored by women are often unsupported, creating barriers to entry for woman newcomers. In this paper, we use the GenderMag method to evaluate GitHub to find cognitive style-specific inclusivity bugs. We redesigned the ``buggy'' GitHub features through a web browser plugin, which we evaluated through a between-subjects experiment (n=75). Our results indicate that the changes to the interface improve users' performance and self-efficacy, mainly for individuals with cognitive styles more common to women. Our results can inspire designers of social coding platforms and software engineering tools to produce more inclusive development environments. 

    % Social coding platforms such as GitHub 
    % allows software developers to work collaboratively. However, if such platforms do not support specific cognitive styles, they disproportionately impact newcomers with these styles, especially those from underrepresented groups such as women. Research shows that the cognitive styles that are common to underrepresented communities are often unsupported, creating an additional barrier to entry for these communities. In this paper, we use the GenderMag method to evaluate GitHub to find cognitive style-specific usability bugs. We redesigned the ``buggy'' GitHub features through a web browser plugin, which we evaluated through a between-subjects experiment (n=75). Our results indicate that the changes to the interface result in an improvement in the performance and self-efficacy of users, mainly for individuals with cognitive styles that are more common to women. Designers of social coding platforms and software engineering tools can use our results to design more inclusive development environments. 

\vspace{1mm}
\textit{General Abstract}---Diversity is an important aspect of society. One form of diversity is cognitive diversity---differences in cognitive styles, which helps generate a diversity of thoughts. Unfortunately, software tools often do not support different cognitive styles (e.g., learning styles), disproportionately impacting those whose styles are not supported. These individuals pay a cognitive ``tax'' each time they use the tools. In this work, we found ``inclusivity bugs" in GitHub, a social coding platform. We then redesigned these buggy features and evaluated them with users. Our results show that the redesign makes it easier for the group of individuals whose cognitive styles were unsupported in the original design, with the percentage of completed tasks rising from 67\% to 95\% for this group.

\end{abstract}

\textbf{\textit{keywords: open source, diversity and inclusion, human factors, cognitive styles, human-computer interaction.}}

\section{Introduction}
\label{sec:introduction}

Open Source Software (OSS) projects play an important role in improving inclusion in workforce development, where contributors join projects to learn new skills~\cite{gerosa2021shifting}, showcase their skills~\cite{von2012carrots}, or improve their career paths~\cite{jergensen2011onion}. Successful participation in OSS projects also helps newcomers gain visibility among their peers~\cite{cai2016reputation, riehle2015open}, benefits society by developing a product used by many users~\cite{parra2016making}, and improves their chances of achieving professional success~\cite{greene2016cvexplorer, riehle2015open}.

However, newcomers to OSS face several challenges~\cite{steinmacher2015social}, and these challenges differently affect underrepresented populations, including those whose cognitive styles are ill-supported by the project's information landscape~\cite{padala2020gender, mendez2018open}. The consequences of these challenges to underrepresented populations may include a steeper learning curve, lack of community support, and obstacles to figuring out how to start contributing, all of which add to the diversity imbalance in OSS~\cite{trinkenreich2021women}. Social diversity has been shown to positively affect productivity, teamwork, and quality of contributions~\cite{horwitz2007effects, vasilescu2015gender}. On the other hand, low diversity has unfortunate effects: (i) OSS projects miss out on the benefits of a more expansive set of contributors and the diversity of thought that these potential contributors could bring; (ii) minorities miss out on the learning and experience opportunities that OSS projects provide; and (iii) minorities miss out on job opportunities when recruiters use OSS contributions to make hiring decisions~\cite{marlow2013impression, singer2013mutual}. Although the lack of diversity in OSS has been well-documented for years, there is limited progress in closing this gap~\cite{trinkenreich2021women, ford2017someone, robles2016women}.

Past work~\cite{padala2020gender, mendez2018open} has shown that the way information is provided in OSS projects (e.g., documentation, issue description) benefits certain cognitive styles (e.g., those who learn by tinkering) over others (e.g., process-oriented learners). The information architecture of OSS project pages (e.g., project description pages and descriptions of issues in the issue tracker) usually appeal to those who have high self-efficacy and are motivated by individual pursuits such as intellectual stimulation, competition, and learning technology for fun. According to Burnett et al.~\cite{burnett2010gender}, these pursuits cater to characteristics associated with men, which can neglect women and other contributors who may have different motivations and personal characteristics (see also \cite{cazan2016computer, singh2013role}). 
This lack of support for diverse user characteristics leads to inclusivity bugs~\cite{chatterjee2021aid, guizani2022debug}---software behaviors that disproportionately disadvantage a particular group of users of that software.

%In prior work, we applied the GenderMag Method to find and fix inclusivity bugs in several OSS projects (see more details about GenderMag~\cite{burnett2016gendermag} in Section~\ref{sec:methodology}). While fixing the bugs, we noticed that several inclusivity bugs were crosscutting for all projects and were caused by the GitHub platform itself. Therefore, in this paper, we looked for inclusivity bugs that can further disadvantage users with underrepresented types of cognitive styles in the GitHub platform. 
%%%OLD
%In prior work, researchers applied the GenderMag method to find and fix inclusivity bugs in several OSS projects (see more details about the GenderMag method~\cite{burnett2016gendermag} in Section~\ref{sec:methodology}). \italo{An inclusivity bug is a software behavior that disproportionately disadvantages a particular group of users of that software~\cite{chatterjee2021aid}.} These fixes, however, were confined to the information architecture of the projects. Several barriers that newcomers face can be because of the lack of support for varied cognitive styles in the GitHub platform itself. 

%In prior work, researchers applied the GenderMag method to find and fix inclusivity bugs in OSS projects~\cite{burnett2016gendermag}. %(see more details about the GenderMag method in Section~\ref{sec:methodology}).

In our study, we investigate inclusivity bugs in the GitHub platform that affect newcomers to this platform. Inclusivity bugs in the platform can have far-reaching impacts on thousands of OSS projects (as of today, more than 200 Million repositories are hosted on GitHub). The following research questions guided our investigation:

\rqone{}

\rqtwo{}

We analyzed four tasks newcomers often perform to make their first pull request on GitHub and found inclusivity bugs in all of them. We redesigned the impacted interface to address the identified bugs and implemented a browser plugin to change the platform interface based on our redesign (we do not have access to change GitHub itself). We evaluated the original and the redesigned interface through a between-subject user study with 75 participants.

%Our contributions include (i) a set of inclusivity bugs that newcomers face when starting to use GitHub, (ii) a GitHub plugin to aid newcomers with different cognitive styles, and (iii) insights into how newcomers' performance can be improved when their cognitive styles are supported.

Our main goal is to mitigate cognitive barriers newcomers face due to inclusivity bugs. As we show in this paper, GitHub, a platform newcomers use to contribute to OSS, creates barriers for users with different characteristics, disproportionately impacting those from underrepresented groups. These barriers may discourage newcomers and add to the existing diversity gaps, as these tools and infrastructure are the main channels through which OSS newcomers interact with the community. This paper provides insights into how newcomers' performance can be improved when their cognitive styles are supported. Providing adequate support for diverse cognitive styles can help improve the overall community diversity.

%To answer these research questions, we developed a plugin that makes changes in the GitHub user interface based on the inclusivity bugs found using the GenderMag method~\cite{burnett2016finding}, that is, an inspection method for software professionals to use to find gender-inclusiveness issues. We designed an experiment to evaluate how well newcomers would perform using the plugin developed to help different cognitive styles using GitHub and compared the results with the group that did not use the approach proposed. 

%The remainder of this paper is structured as follows. Section~\ref{sec:relatedwork} discusses related studies. Section~\ref{sec:methodology} describes the methodology applied in this study. Section~\ref{sec:resultsanddiscussion} provides the research questions answers and discusses the main findings. Section \ref{sec:implications} highlights the study implications. Section \ref{sec:threatstovalidity} presents the threats to validity of this study. Finally, Section~\ref{sec:conclusion} concludes this work and presents some direction for future work.

\section{Related Work}
\label{sec:relatedwork}

This section discusses work related to newcomers' onboarding in OSS, diversity and bias in OSS, and cognitive styles.

%Newcomers Onboarding
\textbf{Newcomer's Onboarding:} Previous work has investigated OSS contribution challenges~\cite{steinmacher2015social, santoshits, hannebauer2017relationship, jensen2011joining, steinmacher2014preliminary}. Steinmacher et al.~\cite{steinmacher2015social} conducted a mixed-method study and identified 58 barriers faced by newcomers. Researchers have also investigated specific types of challenges. For example, toxic environments have been studied in the literature~\cite{bosu2019diversity, prana2021including,guizani2021long}, which evidenced situations in which OSS project members were unfriendly, unhelpful, or elitist~\cite{storey2016social}. Jensen~et al.~\cite{jensen2011joining} analyzed the speed at which emails sent by newcomers are answered, the role played by gender or nationality in the kinds of answers newcomers receive, and the reception newcomers face. A better understanding of the barriers enables communities and researchers to design and produce tools and conceive strategies to better support newcomers~\cite{balali2018newcomers}. Our work complements existing literature by focusing on making social coding platforms more inclusive by supporting the onboarding of newcomers with different cognitive styles.

%Diversity - Women OSS and Bias in OSS
\textbf{Diversity/Bias in OSS:} Low diversity in OSS is a concern raised by different studies in the literature when considering gender~\cite{trinkenreich2021women, guizani2022debug, bosu2019diversity, terrell2016gender}, language~\cite{storey2016social}, and location~\cite{storey2016social}. Past work has shown that diverse teams are more productive~\cite{vasilescu2015gender}. However, minorities face challenges in becoming a part of an OSS community~\cite{trinkenreich2021women}. Most OSS communities function as meritocracies~\cite{feller2000framework}, in which minorities report experiencing ``imposter syndrome''~\cite{vasilescu2015gender}. These competitive settings have been known to discourage minorities such as women in OSS~\cite{miller2012toward, vugt2007gender}. Participant observation of OSS contributors found that ``men monopolize code authorship and simultaneously de-legitimize the kinds of social ties necessary to build mechanisms for women's inclusion''~\cite{nafus2012patches}. Generally, cultures that describe themselves as meritocracies tend to be male-dominated ones that women experience as unfriendly~\cite{turkle2005second}. %The GenderMag method~\cite{burnett2016gendermag} was developed to find and fix inclusivity bugs, reducing the challenges faced by minorities with cognitive styles common to women. It provides multiple lenses to consider five dimensions of cognitive styles, each backed by extensive research~\cite{burnett2016gendermag, stumpf2020gender}. \citet{padala2020gender} employed this method to investigate a variety of OSS projects and found that the project landscape is complicit in creating gender-biased contribution barriers. 
In our work, we aim to reduce the bias found in social coding platforms used by a wide range of users to support them regarding their different cognitive styles to interact with OSS projects.

\textbf{Cognitive styles:} Research has shown that developers have different cognitive styles~\cite{burnett2016gendermag} and motivation~\cite{gerosa2021shifting}, and that cognition plays an essential role in software engineering activities~\cite{fagerholm2022cognition}. For example, more women are task-oriented, whereas more men are motivated to learn a new technology for fun~\cite{padala2020gender, mendez2018open, mendez2018gender}. %Research has shown that women are more statistically likely to use comprehensive information processing styles (e.g., gathering fairly complete information before proceeding). In contrast, men are more statistically likely to use particular styles (e.g., following the first promising information and then backtracking to find a solution)~\cite{coursaris2008empirical, meyers2015revisiting, riedl2010there}. Research has also found that women are statistically less likely to playfully experiment (``tinker'') with new technology as compared to men~\cite{burnett2010gender, cao2010debugging, chang2014specialization, hou2006girls}. 
These differences in cognitive styles may negatively impact how women and men contribute to OSS, and it mainly happens when OSS projects and the underlying infrastructure support certain cognitive styles (e.g., selective information processing or learning by tinkering) and impede others (e.g., comprehensive information processing or process-oriented learning). Our work considers a variety of cognitive styles to propose changes to GitHub to support diverse newcomers.

\section{Research Method}
\label{sec:methodology}

\begin{figure}[!ht]
  \centering
  \includegraphics[width=8cm]{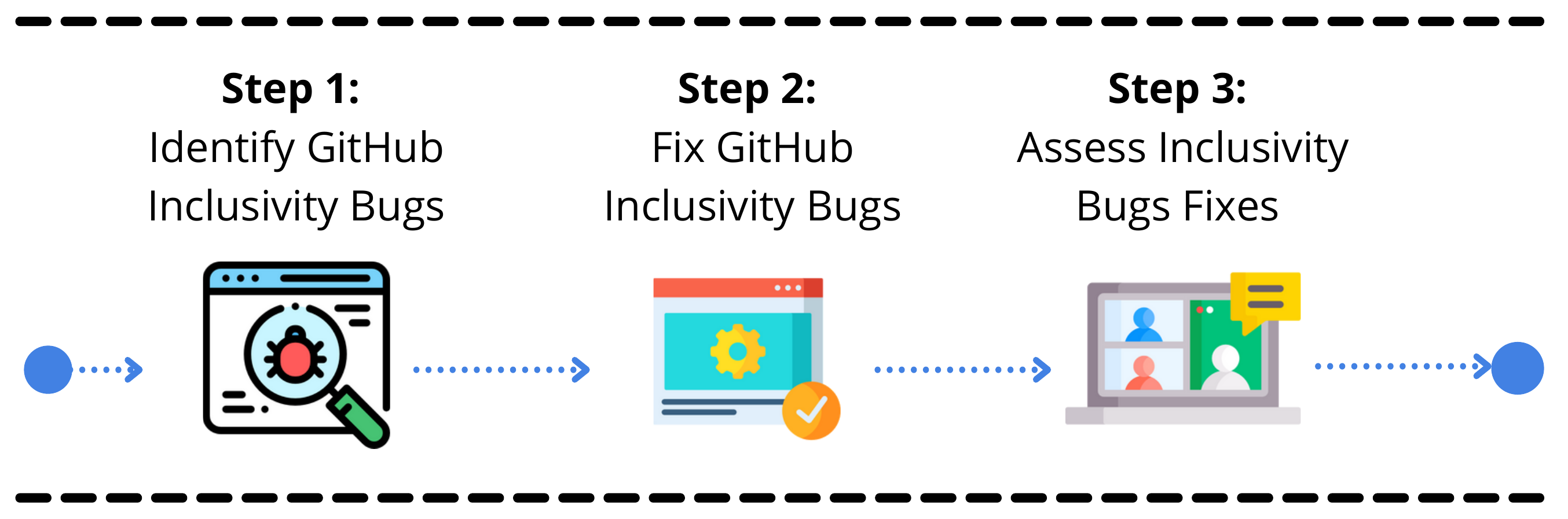}
  \vspace{-2.5mm}
  \caption{Research method overview.}
  \label{fig:methodologyactivity}
\end{figure}

We followed a three-step method, as illustrated in Figure~\ref{fig:methodologyactivity}: (1) we conducted a GenderMag analysis, which has been extensively used to detect gender biases in commercial and OSS products~\cite{mendez2018open, burnett2016finding, cunningham2016supporting, shekhar2018cognitive, vorvoreanu2019gender}; (2) we proposed fixes to the GitHub-related inclusivity bugs and developed a browser plugin to implement these changes in the GitHub interface; and (3) we conducted an experiment to compare the original GitHub interface with the interface enriched by the plugin.

%adapted from the study of Mihaela et al.~\cite{vorvoreanu2019gender}. 
%our research group composed of 4 experienced researchers with more then 10 years of experience and expertise in HCI, SE and design of software systems and 1 student from graduate level and 1 student from undergrad

\vspace{-2.5mm}
\subsection{Step 1 - Identifying GitHub Inclusivity Bugs}\label{sec:gendermag}

To identify inclusivity bugs, we used GenderMag~\cite{burnett2016gendermag}, a systematic inspection method tool builders can use to evaluate their software for inclusivity bugs. GenderMag is based on research showing that individual differences in cognitive styles (referred to as facets) cluster by gender.
The method encapsulates these facets into personas--Abi, Pat, and Tim. Abi and Tim occupy the opposite spectrum of facet values, with the Abi persona aligned with facet values that women tend to favor and Tim embodying facet values typically favored by men. The Pat persona includes a mix of these facet values.

The five facets that GenderMag uses are: (i) \textbf{Motivation:} Abis are motivated to use technology for what they can accomplish with it, whereas Tims are often motivated by their enjoyment of technology per se~\cite{burnett2010gender, margolis2002unlocking, burnett2011gender}; (ii) \textbf{Information processing styles:} Abis process new information comprehensively—gathering fairly complete information before proceeding—but Tims use selective information processing—following the first promising information, then backtracking if needed~\cite{riedl2010there, meyers2015revisiting}; (iii) \textbf{Computer self-efficacy:} relates with a person's confidence about succeeding at a specific task, which influences their use of cognitive strategies, persistence, and strategies for coping with obstacles. Abis have lower computer self-efficacy as compared to their peers; (iv) \textbf{Risk aversion:} Abis are risk-averse when trying out new features as compared to Tims~\cite{dohmen2011individual, charness2012strong}, which impact their decisions about which feature sets to use; and (v) \textbf{Learning: by Process vs. by Tinkering:} Abis prefer process-oriented learning, whereas Tims like to playfully experiment (``tinker'') with software features new to them~\cite{burnett2010gender, beckwith2006tinkering, cao2010debugging}. Each cognitive style has advantages, but either is at a disadvantage when not supported by the software~\cite{chatterjee2022inclusivity}. 

%For example, a study with men and women using a search product showed that women's action failure rates were over twice as high as men's. However, after the product owners fixed the gender-inclusivity bugs GenderMag revealed using the Abi and Tim personas, failure rates of both the participating genders went down, and the difference between these two genders' failure rates completely disappeared~\cite{vorvoreanu2019gender}. 

%\begin{figure}[!ht]
%  \centering
%  \includegraphics[width=\linewidth]{images/abbyfacet.pdf}
%  \vspace{-2.5mm}
%  \caption{Abi persona facets overview~\cite{burnett2018gendermag}.}
%  \label{fig:abbypersona}
%\end{figure}

GenderMag is used by evaluation teams to walk through a use case in the project they are evaluating using Abi, Pat, or Tim personas. At each step of the walkthrough, the team writes down the answers to three questions: 

\begin{itemize}
    \item \textbf{SubgoalQ:} Will Abi have formed this subgoal as a step to their overall goal? (Yes/no/maybe, why, facets involved).
    
    \item \textbf{\textit{ActionQ1:}} Will Abi know what to do at this step? (Yes/no/maybe, why, facets involved).
    
    \item \textbf{\textit{ActionQ2:}} If Abi does the right thing, will s/he know s/he did the right thing and is making progress toward their goal? (Yes/no/maybe, why, facets involved).
\end{itemize}

When the answer to any of these questions is negative, the team identifies a potential bug; if the ``why'' relates to a particular cognitive style, this shows a disproportionate effect on people who have that cognitive style---i.e., an \textit{inclusivity bug}. Thus, a team's answers to these questions become their inclusivity bug report, which they can process and prioritize the same way they would with any other type of bug report.

We selected the Abi and Tim personas~\cite{burnett2018gendermag} as they represent opposite ends of the GenderMag facet ranges.
We customized the persona's profile to represent our target users: newcomers looking to make their first contribution using GitHub and have never performed a pull request (PR) before. We identified four use cases (i.e., edit a file, submit a pull request, fork repository, upload a new file) as described in Table~\ref{tab:gendermagscen}.

% We defined a set of contribution goals and subgoals, as described in . \italo{
% These goals were chosen because they are often part of the process to make the first contribution to an OSS project~\cite{steinmacher2016overcoming}.} 

% \red{In this paper, we differentiate tasks from goal. A task refers to the actions performed by the participants in the experiment, while goals refers to how we organized the GenderMag analysis. While these terms are different from each other, they have a close association (e.g., Task 1 is associated with Goal 1).}

\begin{table}[!ht]\scriptsize
\centering
\vspace{-2.5mm}
\caption{GenderMag analysis use case and subgoals}
\vspace{-2mm}
\label{tab:gendermagscen}
\begin{tabular}{m{32mm}|m{43mm}}
\hline
\multicolumn{1}{c|}{\textbf{Use Case}} & \multicolumn{1}{c}{\textbf{Subgoals}} \\ \hline \hline
\multirow{2}{*}{\parbox{32mm}{UC\#1 - Submit a pull request}} & \#1.1 - Make a change to a README file  \\ \cline{2-2} 
 & \#1.2 - Submit the pull request \\ \hline
 
UC\#2 - View changed files in PR & \#2.1 - Find the changed files in the interface \\ \hline

UC\#3 - Request help to solve the pull request  & \#3.1 - Find an experienced contributor in the project to ask for help to solve the PR \\ \hline
 
\multirow{2}{*}{\parbox{32mm}{UC\#4 - Upload file}} & \#4.1 - Discover how to upload a file  \\ \cline{2-2} 

 & \#4.2 - Request push access to upload file  \\ \hline \hline
\end{tabular}
\end{table}

Given these personas and use cases, 6 members of our research group conducted the GenderMag walkthroughs on GitHub-hosted projects using the procedures defined by Burnett et al.~\cite{burnett2018gendermag}. The group had prior training and experience in conducting GenderMag analysis. As a first step, the group identified the subgoals and actions for each use case. We then performed the GenderMag evaluations for each use case by first using the Abi persona and then another set of evaluations with the Tim persona. We identified 12 inclusivity bugs in different parts of the GitHub interface. 

%We did one walkthrough of each goal from the perspective of the Abi persona and separate walkthroughs of the same goals with the Tim persona. 

%Then, we conducted a GenderMag walkthrough focused on the Abi persona to identify inclusiveness bugs. 

%The definitions of our subgoals are related to the features that a newcomer would have to use to make a successful contribution to an OSS project .

%We investigated in the context of GitHub\footnote{https://github.com/}, which is a hosting site where developers and programmers can upload the code they create and work collaboratively to develop new features.

%According to the GenderMag analysis~\cite{burnett2016gendermag}, goals must be broken down into subgoals. Subgoals provide digestible ``abstract'' sequences through the goal. A subgoal has concrete action sequences users should carry out. GenderMag analysis asks whether users would want and be able to perform these actions.

%\end{comment}

\subsection{Step 2 - Fixing GitHub Inclusivity Bugs}
%can be fixed in multiple ways

%As stated by~\citet{vorvoreanu2019gender}, the GenderMag analysis outcomes can identify not only inclusivity bugs but why the bug might arise and what specific problem-solving facet(s) are implicated. Therefore, we used the results from the GenderMag analysis to redesign the GitHub interface to address the inclusivity bugs identified in Step 1. As an example for Goal \#1, we identified an issue in Abi's learning process and self-efficacy facet that would affect her ability to edit a file in an OSS project (Figure~\ref{fig:gitoriginterface}). \italo{Users' self-efficacy can be applied to many contexts and can affect an individual's ultimate success regarding whether they blame themselves when something goes wrong. The individual's self-efficacy can also affect their willingness to continue solving a problem in the face of difficulty and whether they approach another angle if their first attempt fails~\cite{bandura1977self}.} Figuring out that the README is among all the files in the OSS project may be difficult for Abi users. 
%The file also appears below the repository files, but the user needs to scroll down.

\begin{comment}
\begin{figure}[!ht]
  \centering
  \includegraphics[width=8.5cm]{images/gitinterface1.pdf}
  \vspace{-1.7mm}
  \caption{GitHub original interface.}
  \label{fig:gitoriginterface}
\end{figure}
\end{comment}

We redesigned the GitHub interface to support the GenderMag facets that were previously unsupported and caused the inclusivity bugs we identified in Step 1. As stated by Guizani et al.~\cite{guizani2022debug}, the outcomes of GenderMag analysis point not only to inclusivity bugs but also to why the bugs might arise and what specific problem-solving facet(s) are implicated. 

\begin{figure}[!h]
  \centering
  \vspace{-1.7mm}
  \includegraphics[width=8cm]{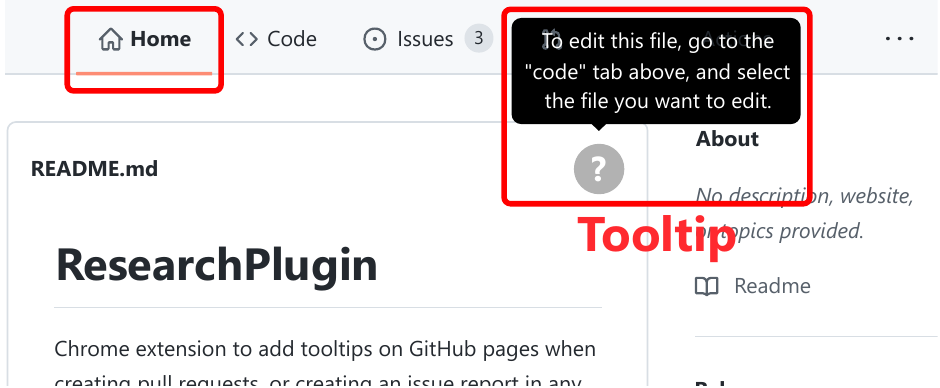}
  %\vspace{-1.7mm}
  \caption{GitHub interface modified by the developed plugin.}
  \label{fig:gitoriginterfacepost}
\end{figure}

As an example of redesign, for UC\#1, we identified an issue related to Abi's process-oriented learning style and self-efficacy facets that would affect her ability to edit a file in an OSS project. 
%\italo{Editing a file was part of the use case of submitting a pull request.}
%Users' self-efficacy can be applied to many contexts and can affect an individual's ultimate success regarding whether they blame themselves when something goes wrong. The individual's self-efficacy can also affect their willingness to continue solving a problem in the face of difficulty and whether they approach another angle if their first attempt fails~\cite{bandura1977self}. 
The redesign focused on Abi's process-oriented learning facet to give explicit guidance on submitting a pull request by leveraging the design principle of ``visibility.'' We did so by: (1) presenting the README file information to users more explicitly through a new tab called home (Figure~\ref{fig:gitoriginterfacepost}), which highlights the importance of the README file, and (2) including a tooltip to explain that the user can edit the file: \textit{To edit this file, go to the ``code'' tab above, and select the file you want to edit.} Our proposed solution also addressed Abi's self-efficacy facet by showing that she is on the right track to completing the subgoal (\#1.1, make changes to README). 

Once our research team agreed with the redesign solutions proposed for each issue identified in Step 1, we started the development of a plugin to change GitHub's interface. The plugin was developed as a Chrome extension to change the original GitHub interface. The plugin is developed in JavaScript and uses the GitHub API to collect data about a user in JSON format. It is available on GitHub\footnote{https://github.com/NAU-OSL/ResearchPlugin} for anyone interested in using it and making contributions, as well as in the supplementary material\footnote{\url{https://figshare.com/s/4e7724bde0b1d47ecaeb}}.
%\footnote{GitHub link is redacted due to double-blind}.

\subsection{Step 3 - Assessing the Inclusivity Bug Fixes}\label{sec:assessing-inclusivity}

Finally, we conducted an experiment to evaluate how the modified interface changed the user experience for Abis. Even though inclusivity bugs can be fixed in multiple ways, we expected that we would reduce an eventual performance gap between Abis and Tims who use the modified interface, since the modifications were supported by the analytic/theory-based method.

We follow the guidelines provided in~\cite{wohlin2012experimentation} to report our experiment. We conducted an experiment to \textit{analyze} how the proposed plugin supports newcomers with different cognitive styles. We compared users using GitHub's original version to users using our GitHub plugin, \textit{{for the purpose of}} evaluating, \textit{{with respect to their}} effectiveness in completing the use cases, \textit{{from the point of view of the}} researchers, \textit{{in the context of}} the GitHub environment when a newcomer attempts to make their first contribution.

The participants interacted with a copy of a community-based OSS project named JabRef\footnote{\url{https://github.com/JabRef/jabref}}. Participants completed the four use cases used for finding the bugs (Table~\ref{tab:gendermagscen}): 

\begin{itemize}
\item UC\#1 - Submit a pull request: The newcomer needs to edit a file in the project and submit the changes via a pull request (PR); 
\item UC\#2 - View changed file. In this task, we asked participants to analyze an open pull request and find which files were changed when this pull request was created; 
\item UC\#3  - Request help to solve the PR. The participant needs to find an experienced project contributor and invite them to work together to solve the pull request; and 
\item UC\#4 - Upload a file: The participant should try to upload a new file to the repository. 
\end{itemize}

We conducted a pilot study with five researchers outside our group to collect feedback about the instruments (questionnaires and use case definitions) and study design. The pilot study helped to improve our instruments. We used an iterative process to apply the necessary changes after each pilot session. This resulted in more detailed scripts and documentation about the use cases. We ran new pilot sessions until we reached a consensus that the instruments were reliable enough to start the actual study. A replication package with this material is available online (see the previous subsection). The replication package also includes the developed GitHub plugin and installation instructions.

%The study was initially conducted with one participant for a session with a facilitator and an observer, with 37 sessions conducted individually. However, to optimize the study data collection, we asked professors to apply the study during their classes, where the participants had to follow the instructions provided and perform the study. The participants were in the same room and could ask questions to the professor if they needed any help or did not understand the required tasks. We collected additional 38 valid responses for the study.

We recruited 75 undergraduate students from diverse STEM majors from 5 distinct universities in the US and Brazil. The majority of participants were pursuing Computer Science majors. Our recruiting criteria were students who knew how to program but had never opened a pull request on GitHub, so previous experiences with the interface would not bias them. We opted to recruit undergraduate students for our study because the literature mentions that educators have been using OSS to train students, and these students are potential OSS project contributors~\cite{steinmacher2016overcoming}. We asked the students if they had previous experience with GitHub and OSS. Some of them responded that they had used GitHub once (Plugin = 10 and Control = 7), but when we questioned about what they had used GitHub for, they said that they just created the account but never contributed to any project, so they fit our criteria (never opened a pull request). We also asked about their experience with OSS, and a few participants answered that they had some experience (Plugin = 4 and Control = 3). When we asked what kind of experience they had, they informed us that they had studied OSS concepts in previous courses in college.

We used a between-subject design to balance participants in the original version (Control group) and GitHub plugin version (Plugin group) by GenderMag facets~\cite{montoya2022selecting, vorvoreanu2019gender}. We used GenderMag's questionnaire to assess participants' facets with 9-point Likert items \cite{anita2023}. We ended up with different numbers of participants between the two treatments, as some of the participants were a no-show, Table~\ref{tab:numberofpart}). 

Unfortunately, we had a small sample of women participants (18 vs. 57) due to the gender distribution of students in the classes we recruited from. We attempted to balance the participants in each treatment based on their cognitive facets, achieving an almost equal distribution of Abis (37) and Tims (38) across the treatment groups. Table~\ref{tab:numberofpart} presents the participants' characteristics in each group.

In the beginning, we conducted each user session one participant at a time with a facilitator and an observer. The participants were asked to perform the four use cases described in Table~\ref{tab:gendermagscen}. We collected audio recordings and observation notes from the sessions and qualitatively analyzed participants' data. We conducted those individual sessions with 50\% of our participants. Then we decided to optimize the data collection by conducting the experiment with students from two classes where we provided an online questionnaire with all the instructions they had to follow to participate in the experiment. A researcher was present the whole time to assist the students in case they needed help or had any questions.

%By opting for this optimized process, we were able to get more quantitative data, however, we did not gather the qualitative data related to the recording of participants' audios.}

%This restriction is essential because we want to see the challenges those newcomers face in this situation. 

We performed a quantitative analysis by collecting the percentage of use cases completed by participants in each group and applied a self-efficacy survey to measure newcomers' confidence in using GitHub. 
%Table~\ref{tab:partdemograph} shows participants' demographic data describing their gender, persona, experience with GitHub and OSS, and which facet each participant has according to their persona. 

%Based on those answers, we characterized the participants' cognitive styles according to the facets of the GenderMag personas Tim and Abi 
%according to (i) the number of participants in each group (control (36) and plugin (39)); (ii) the number of participants divided by cognitive style (Tim (38) and Abi (37)); and (iii) the number of participants divided by gender (man (57) and woman (18)).} 
%As noted, despite Abi's cognitive style being more common to women, some men also have this cognitive style.

% For instance, if a participant has three facets from Tim (e.g., High Self-Efficacy, Learning: Tinkerer, and Risk Tolerant) and only 2 Abi facets (e.g., Motivations: Technology because the task needs it and Comprehensive Information Processing), the user was classified as having a Tim persona. 

%(e.g., Chemistry, Physics, Civil Engineering, Electric Engineering, Data Science, and Computer Science) 

\begin{table}[!ht]\scriptsize
\centering
\vspace{-2.5mm}
\caption{Number of participants in the experiment}
\vspace{-2.5mm}

\label{tab:numberofpart}
\begin{tabular}{cc|cc|cc}
\cline{2-6}
 &  & \multicolumn{2}{c|}{\textbf{Facets}} & \multicolumn{2}{c}{\textbf{Gender}} \\ \cline{3-6} 
 
\multirow{-2}{*}{\textbf{}} & \multirow{-2}{*}{\textbf{Subjects}} & \multicolumn{1}{c|}{\textbf{Tim}} & \textbf{Abi} & \multicolumn{1}{c|}{\textbf{Man}} & \textbf{Woman} \\ \hline \hline

\multicolumn{1}{c|}{\textbf{Control}} & {\textbf{36}} & \multicolumn{1}{c|}{\textbf{18}} & {\textbf{18}} & \multicolumn{1}{c|}{\textbf{30}} & {\textbf{6}} \\ \hline

\multicolumn{1}{c|}{\textbf{Plugin}} & {\textbf{39}} & \multicolumn{1}{c|}{\textbf{20}} & {\textbf{19}} & \multicolumn{1}{c|}{\textbf{27}} & {\textbf{12}} \\ \hline

\multicolumn{1}{c|}{\textbf{Total}} & {\textbf{75}} & \multicolumn{1}{c|}{\textbf{38}} & {\textbf{37}} & \multicolumn{1}{c|}{\textbf{57}} & {\textbf{18}} \\ \hline \hline

\end{tabular}
\end{table}
\vspace{-2mm}

We also administered a questionnaire in which participants provided their self-perception about their ability to complete use cases using GitHub, i.e., self-efficacy to complete specific tasks. The questionnaire was based on the work of Bandura~\cite{bandura2014social} and had 5 items. Participants answered those questions before and after the experiment using a 5-point Likert scale ranging from strongly disagree to strongly agree (with a neutral option). The goal was to capture the students' self-perceived efficacy about the use case before and after they attempted executing it. The items were prefixed with ``I am confident that I can:'' followed by: (i) \ldots use GitHub to contribute to projects; (ii) \ldots open a pull request using the GitHub web interface; (iii) \ldots change a file and submit the changes to the project using GitHub; (iv) \ldots find someone to help me using the GitHub web interface; and (v) \ldots submit a new file to a project using GitHub.

In addition to the quantitative analysis, we qualitatively analyzed participants' comments to the open questions of the survey following open 
%and axial 
coding procedures~\cite{strauss1998basics}. We asked participants after each use case to explain any difficulties they experienced in accomplishing the task and what in the interface helped them. Our goals were to understand (i) students' difficulties in using the original and the modified interfaces; and (ii) what in the interfaces helped students the most to complete each use case. The analysis was performed by two authors and validated by a third author. The analysis took around one month.

For our study, we considered the following variables:
(i)~the \textbf{dependent variables} comprise the successful completion of each use case by the participants (Y/N), and (ii) the \textbf{independent variables} are the use of the Plugin and the GenderMag facets (whether the participant is Tim- or Abi-like).

\begin{comment}
\begin{table}[!ht]\scriptsize
\centering
\vspace{-2.5mm}
\caption{Experiment Variables}
\label{tab:experimentvariables}
\begin{tabular}{l}
\hline
\textbf{Dependent} \\ \hline \hline
\multicolumn{1}{c}{- Use Case completion (UC\# 1, 2, 3 and 4)} \\ \hline
\textbf{Independent} \\ \hline \hline
- Plugin (use or non-use) \\ \hline
- Personas (Tim, Abi) \\ \hline
\end{tabular}
\end{table}
\end{comment}

\section{Results}
\label{sec:results}

%In this section, we present the results of our study.

\begin{table*}[!ht]\scriptsize
\centering
\vspace{-2.5mm}
\caption{GitHub Inclusivity Bugs and Proposed Fixes}
\label{tab:inclusivitybugs}
\newcommand{\pb}[1]{\parbox[t][][t]{1.0\linewidth}{#1} \vspace{-2pt}}
\begin{tabular}{m{16mm}|c|m{35mm}|m{28mm}|m{73mm}}
\hline
 \multicolumn{1}{c|}{\textbf{Use Case}} & \multicolumn{1}{c|}{\textbf{\# Bugs}} & \multicolumn{1}{c|}{\textbf{Bug Description}} & \multicolumn{1}{c|}{\textbf{GenderMag Facets}} & \multicolumn{1}{c}{\textbf{Bug Fixes}} \\ \hline \hline

\multirow{14}{*}{\pb{\centering\#1 \\ Submit pull request}}

 & 1 & Difficulty in finding Readme file to edit; & \pb{- Learning: Process vs. Tinkering;\\ - Computer self-efficacy.} & \pb{- Add ``Home'' link to the navbar to highlight the importance of the Readme File in the repository hierarchy. This link presents this file's content and includes a tooltip to explain that the user can edit the file.} \\ \cline{2-5}
 & 2 & \pb{After clicking to edit the file, difficulty in finding the options to edit the file;} & \pb{- Learning: Process vs. Tinkering;\\ - Computer self-efficacy;\\ - Attitude Towards Risk.} & \pb{- Include a progress bar, to indicate the steps of the workflow related to this task;\\ - Include a tooltip to explain what happens in case the user changes the original filename.} \\ \cline{2-5} 
 & 3 & Difficulty in understanding the commit form; & \pb{- Learning: Process vs. Tinkering;\\ - Computer self-efficacy;\\ - Attitude Towards Risk.} & \pb{- Put tooltips and field labels explaining form fields to help the user understand the importance of informing a commit message.} \\ \cline{2-5} 
 & 4 & \pb{Difficulty in understanding the workflow after the file is edited;} & \pb{- Motivations;\\ - Learning: Process vs. Tinkering;\\ - Information Processing Style.} & \pb{- We have the progress bar, to indicate that this step is important to complete the task;\\ - Include a tooltip to explain the conflict message that appears;\\ - Include a tooltip to explain the code that is related to the changes.} \\ \cline{2-5} 
 & 5 & \pb{Lack of feedback indicating if the creation of the pull request was successful;} & \pb{- Learning: Process vs. Tinkering;\\ - Computer Self-Efficacy;\\ - Information Processing Style.} & \pb{- After the click on the create pull request button, redirect to a page with a success message and a progress bar showing that the pull request is completed;\\ - Include a tooltip to explain what does the ``Close pull request'' button do.} \\ \hline \hline
 
 \multirow{1}{*}{\pb{\centering\#2 View \\changed files}}
 
 & 6 & \pb{Difficulty in understanding the workflow after the user opens a pull request;} & \pb{- Learning: Process vs. Tinkering;\\ - Computer Self-Efficacy.} & \pb{- Include a tooltip to highlight the navbar that describes some actions that can be made in the pull request.} \\ \hline \hline
 
\multirow{3}{*}{\pb{\centering\#3 \\ Request help to solve the PR}} 
 
 & 7 & \pb{Difficulty in finding the option to mention another contributor;} & \pb{- Learning: Process vs. Tinkering;\\ - Computer Self-Efficacy.} & \pb{- Include a tooltip in the @ symbol icon to say "Use it to mention a contributor."} \\ \cline{2-5} 
 & 8 & \pb{Lack of feedback about the action of mentioning another contributor;} & \pb{- Information Processing Style;\\ - Computer Self-Efficacy.} & \pb{- Add a confirmation message to let the user know that the mentioned contributor will receive a notification and may help the newcomer in this issue.} \\ \hline \hline
 
\multirow{7}{*}{\pb{\centering\#4 \\ Upload file}} & 9 & \pb{Difficulty in understanding the steps needed to upload a file;} & \pb{- Information Processing Style;\\ - Attitude Towards Risk;\\ - Motivations.} & \pb{- Change the message to inform that it is necessary to fork; \\ - Make the fork button green to highlight that it is enabled.} \\ \cline{2-5} 
 & 10 & \pb{Lack of feedback indicating if the action of forking the repository is completed;} & \pb{- Information Processing Style;\\ - Computer Self-Efficacy.} & \pb{- Add a success message to the page that appears after the click on the fork button.} \\ \cline{2-5} 
 & 11 & \pb{Difficulty in understanding the commit form;} & \pb{- Learning: Process vs. Tinkering;\\ - Motivations.} & \pb{- Include tooltips explaining the form fields to make the newcomer understand the importance of informing a commit message.} \\ \cline{2-5} 
 & 12 & \pb{Lack of feedback indicating if the action of uploading the file is completed;} & \pb{- Information Processing Style;\\ - Learning: Process vs. Tinkering;} & \pb{- Add a success message to the repository page that appears after the click on the commits changes button.} \\ \hline \hline
\end{tabular}
\end{table*}

\subsection{Discovering and Fixing inclusivity bugs on GitHub}
%\rqone{}

We answer RQ1 based on the results of the GenderMag evaluation of the GitHub interface, which uncovered 12 inclusivity bugs. Table~\ref{tab:inclusivitybugs} summarizes the inclusivity bugs, associated GenderMag facets, and how we fixed them. The fixes leveraged the design principles of visibility and feedback, along with the tenet of clarity of instructions and reduction of information load where appropriate. The specific UI design changes were inspired by successful fixes to inclusivity bugs as compiled in the GenderMag design catalog\footnote{\url{https://gendermag.org/dc/}}. The parts of the GitHub interface where these bugs were found can be accessed in the supplementary material\footnote{\url{https://figshare.com/s/4e7724bde0b1d47ecaeb}}. 

\begin{figure}[!ht]
  \centering
  \vspace{-1.7mm}
  \includegraphics[width=8cm]{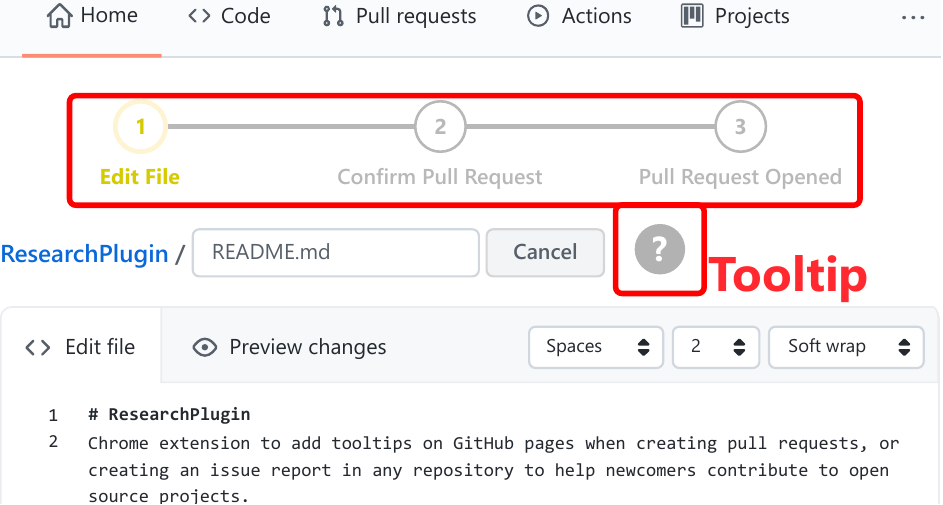}
  \caption{UC\#1 / Bugfix \#2 - plugin interface: inclusion of progress bar and tooltip.}
  \vspace{-3mm}
  \label{fig:commit}
\end{figure}

In UC\#1 - Submit pull request, we investigated the GitHub interface that an average user interacts with to edit a file and open a pull request. This use case involved five inclusivity bugs. Among the reported bugs, we found Abi would have difficulty understanding the workflow (what to do next) after the file was edited (Bug \#4). Abis are comprehensive information processors and process-oriented learners; in this interface, they would not have all the information needed to complete the task and are unlikely to tinker to figure out how to complete it. 
To address this bug, we proposed: (1) A progress bar indicating the steps of the workflow (improved feedback), allowing Abis to know upfront the process needed to complete the use case; and (2) a tooltip (improved visibility) to explain what happens when the file is edited to provide additional information if Abis need it (Figure~\ref{fig:commit}). Instructions as a tooltip reduce clutter and do not disadvantage tinkerers like Tim.

\begin{figure}[!ht]
  \centering
  \includegraphics[width=8cm]{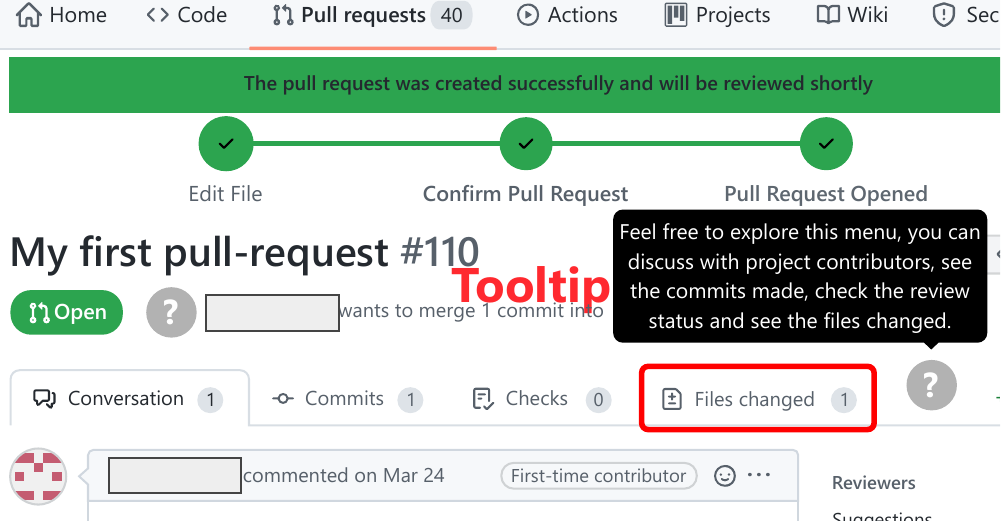}
  \vspace{-1.7mm}
  \caption{UC\#2 / Bugfix \#6 - plugin interface: inclusion of tooltip to guide users.}
  \label{fig:fileschanged}
\end{figure}

UC\#2, View changed files, included one inclusivity bug (Bug \#6), where Abi has difficulty understanding what to do next after opening the pull request. In GitHub, after a user opens a pull request, they are directed to a different page, which does not inform what can be done on that page. On reaching this page, Abis, who are process-oriented learners with lower self-efficacy, would be lost, not knowing what to do next. They would not know if they were progressing towards their goal and would be unlikely to tinker around to figure out how to close the pull request.
Our solution adds a tooltip to the navbar that describes some actions that can be made on the pull request page (improved visibility) (Figure~\ref{fig:fileschanged}).

\begin{figure}[!ht]
  \centering
  \includegraphics[height=3.3cm]{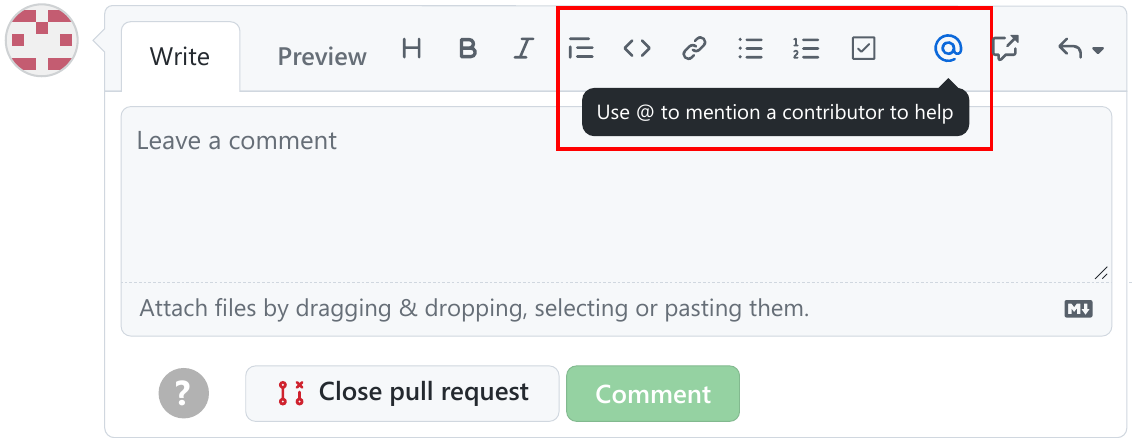}
  \vspace{-1.7mm}
  \caption{UC\#3 / Bugfix \#7 - plugin interface: inclusion of tooltip to guide users.}
  \label{fig:comentario}
  \vspace{-3mm}
\end{figure}

In UC\#3 - Request help to solve the PR, we found 2 inclusivity bugs that could affect users' performance with Abi's cognitive style. The pull request interface is not straightforward. Once the user opens the pull request, it is not clear that it is possible to mention someone in the comment box to ask for help. This lack of information affects users with Abi's facets of learning by process and computer self-efficacy. To address this bug, we included a tooltip in the @ symbol icon to display ``\textit{Use @ to mention a contributor to help},'' as illustrated in Figure~\ref{fig:comentario}. 

\begin{figure}[!ht]
  \centering
  \includegraphics[width=8cm]{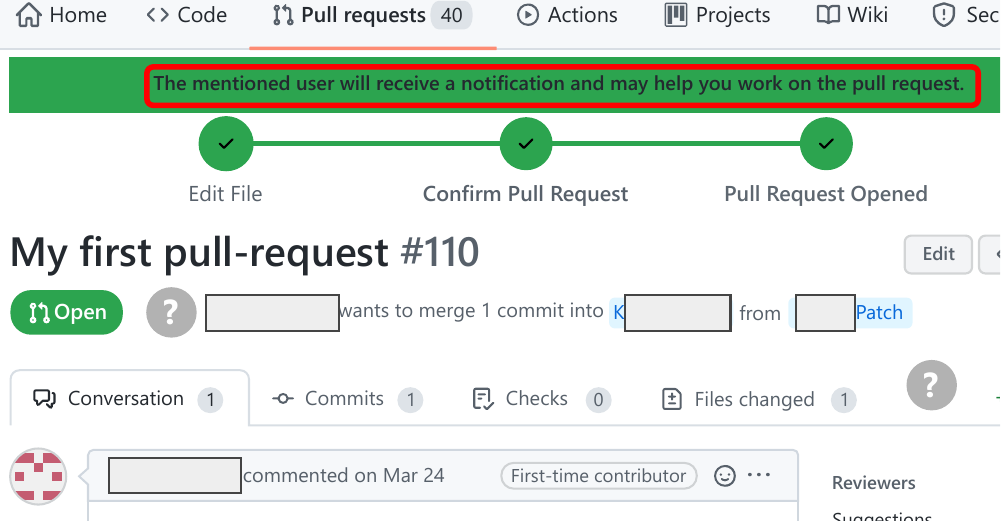}
  %\vspace{-1.7mm}
  \caption{UC\#3 / Bugfix \#8 - plugin interface: inclusion of confirmation message to provide feedback to users.}
  \label{fig:message}
  \vspace{-2.5mm}
\end{figure}

Moreover, after the mention is made, the GitHub interface does not give any feedback about what happens next, affecting comprehensive information processors such as Abi. This can impair Abis' ability to continue with the pull request given their lower self-efficacy, where such users are likely to blame themselves and quit. Even if they asked for help, they would be unsure if the mentioned developer would receive a notification to help them. To fix this bug, we proposed the addition of a confirmation message (improved feedback) to the top of the page informing that: \textit{The mentioned user will receive a notification and may help you to work on the pull request}, as illustrated in Figure~\ref{fig:message}.

\begin{figure}[!ht]
  \centering
  \includegraphics[width=7cm]{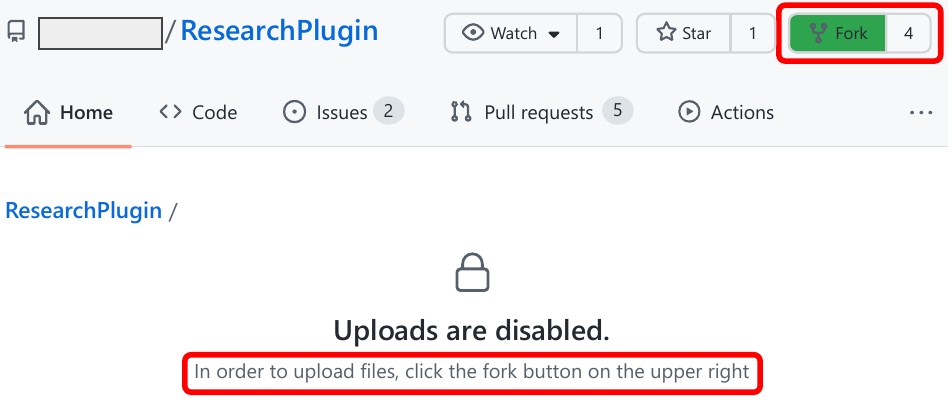}
  \vspace{-1.7mm}
  \caption{UC\#4 / Bugfix \#9 - plugin interface: change of message and color of the fork button.}
  \label{fig:forktwo}
\end{figure}

In UC\#4 - Upload a file, to upload an image to an OSS project, the user needs to have push access to it. For this use case, we found 4 inclusivity bugs. The major bug is related to the second subgoal: it is not possible to upload a file because the newcomer does not have a repository fork nor push access to the original repository. The interface only presents the message that the user needs to have push access to the repository but no direction about how to do it. This bug impacts Abi's facets of comprehensive information processing style, risk averseness, and task-oriented motivations. %\italo{We want to highlight that the motivations cognitive facet spectrum ranges from task-oriented to tech-oriented and is defined by an individual's goal while problem-solving using technology~\cite{burnett2016gendermag}. According to~\citet{hou2006girls}, task-oriented individuals use technology to accomplish their task at hand, while tech-oriented individuals tend to dive into the complexities and try to understand the underlying functionality while using technology.} 
We proposed the following fixes to address this bug: we changed the message to give better feedback informing the user that it is necessary to fork the repository and made the fork button green to highlight that it is enabled on the page. The new message states \textit{In order to upload files, click the fork button in the upper right} (see Figure~\ref{fig:forktwo}).

\rqone[
    \tcblower
    \textbf{Answer:} We found 12 inclusivity bugs after applying the GenderMag inspection method in four use cases a newcomer may perform. These bugs are generally correlated with two or more GenderMag facets. We used the principles of improving visibility, feedback, and reducing information overload in designing the fixes. 
]{}

%-----------------------------------------------------------------
\subsection{Effects of removing GitHub inclusivity bugs}
%\rqtwo{}

\textit{Impact on completion rates.} In RQ2, we investigate how our redesign in Step 1 impacted Abis and Tims. Table~\ref{tab:matrixcompleted} presents the number of participants who correctly completed the tasks, comparing the treatment groups and the different persona facets.
We evaluated the effectiveness of both groups in completing the tasks using the \textit{Chi-Square test} to check the independent relationship between the two categorical variables~\cite{sureiman2013conceptual} (see Table \ref{tab:difference}). 

\begin{table*}[!bp]\scriptsize
\centering
\vspace{-2.5mm}
\caption{Number of tasks completed or failed by participants}
\vspace{-2mm}

\label{tab:matrixcompleted}
\begin{tabular}{lcc|cc|cc|cc|cc}
\cline{2-11}
 & \multicolumn{2}{c|}{\textbf{UC\#1}} & \multicolumn{2}{c|}{\textbf{UC\#2}} & \multicolumn{2}{c|}{\textbf{UC\#3}} & \multicolumn{2}{c|}{\textbf{UC\#4}} & \multicolumn{2}{c}{\textbf{All Use Cases}} \\ \cline{2-11}
 
\multicolumn{1}{c}{\textbf{}} & \multicolumn{1}{c|}{\textbf{Completed}} & \textbf{Failed} & \multicolumn{1}{c|}{\textbf{Completed}} & \textbf{Failed} & \multicolumn{1}{c|}{\textbf{Completed}} & \textbf{Failed} & \multicolumn{1}{c|}{\textbf{Completed}} & \textbf{Failed} & \multicolumn{1}{c|}{\textbf{Completed}} & \textbf{Failed} \\ \hline \hline

\multicolumn{1}{l|}{\textbf{Control}} & \multicolumn{1}{c|}{\cellcolor[HTML]{65C194}33} & \cellcolor[HTML]{FDF3F3}3 & \multicolumn{1}{c|}{\cellcolor[HTML]{61BF91}34} & \cellcolor[HTML]{FEF7F7}2 & \multicolumn{1}{c|}{\cellcolor[HTML]{7DCBA4}28} & \cellcolor[HTML]{F9DDDD}8 & \multicolumn{1}{c|}{\cellcolor[HTML]{CCEBDC}11} & \cellcolor[HTML]{EA9595}25 & \multicolumn{1}{c|}{\cellcolor[HTML]{84CDA9}106} & \cellcolor[HTML]{F7D7D7}38 \\ \hline

\multicolumn{1}{l|}{\textbf{Plugin}} & \multicolumn{1}{c|}{\cellcolor[HTML]{5CBD8D}38} & \cellcolor[HTML]{FFFCFC}1 & \multicolumn{1}{c|}{\cellcolor[HTML]{57BB8A}39} & \cellcolor[HTML]{FFFFFF}0 & \multicolumn{1}{c|}{\cellcolor[HTML]{60BF90}37} & \cellcolor[HTML]{FEF8F8}2 & \multicolumn{1}{c|}{\cellcolor[HTML]{60BF90}37} & \cellcolor[HTML]{FEF8F8}2 & \multicolumn{1}{c|}{\cellcolor[HTML]{5DBE8E}151} & \cellcolor[HTML]{FFFBFB}5 \\ \hline

\multicolumn{1}{l|}{\textbf{Abi - Control}} & \multicolumn{1}{c|}{\cellcolor[HTML]{77C8A0}15} & \cellcolor[HTML]{FAE3E3}3 & \multicolumn{1}{c|}{\cellcolor[HTML]{62C092}17} & \cellcolor[HTML]{FEF6F6}1 & \multicolumn{1}{c|}{\cellcolor[HTML]{96D5B6}11} & \cellcolor[HTML]{F4C6C6}7 & \multicolumn{1}{c|}{\cellcolor[HTML]{CBEADB}5} & \cellcolor[HTML]{EA9696}13 & \multicolumn{1}{c|}{\cellcolor[HTML]{8FD2B1}48} & \cellcolor[HTML]{F5CDCD}24 \\ \hline

\multicolumn{1}{l|}{\textbf{Tim - Control}} & \multicolumn{1}{c|}{\cellcolor[HTML]{57BB8A}18} & \cellcolor[HTML]{FFFFFF}0 & \multicolumn{1}{c|}{\cellcolor[HTML]{60BF90}17} & \cellcolor[HTML]{FEF8F8}1 & \multicolumn{1}{c|}{\cellcolor[HTML]{68C296}17} & \cellcolor[HTML]{FCF0F0}1 & \multicolumn{1}{c|}{\cellcolor[HTML]{CDEBDC}6} & \cellcolor[HTML]{EA9494}12 & \multicolumn{1}{c|}{\cellcolor[HTML]{7BCAA3}58} & \cellcolor[HTML]{F9DFDF}14 \\ \hline 

\multicolumn{1}{l|}{\textbf{Abi - Plugin}} & \multicolumn{1}{c|}{\cellcolor[HTML]{57BB8A}18} & \cellcolor[HTML]{FFFFFF}1 & \multicolumn{1}{c|}{\cellcolor[HTML]{57BB8A}19} & \cellcolor[HTML]{FFFFFF}0 & \multicolumn{1}{c|}{\cellcolor[HTML]{63C093}18} & \cellcolor[HTML]{FDF5F5}1 & \multicolumn{1}{c|}{\cellcolor[HTML]{6FC59B}17} & \cellcolor[HTML]{FBEAEA}2 & \multicolumn{1}{c|}{\cellcolor[HTML]{5BBD8D}72} & \cellcolor[HTML]{FEF7F7}4 \\ \hline

\multicolumn{1}{l|}{\textbf{Tim - Plugin}} & \multicolumn{1}{c|}{\cellcolor[HTML]{5EBE8F}20} & \cellcolor[HTML]{FEF9F9}0 & \multicolumn{1}{c|}{\cellcolor[HTML]{57BB8A}20} & \cellcolor[HTML]{FFFFFF}0 & \multicolumn{1}{c|}{\cellcolor[HTML]{5EBE8F}19} & \cellcolor[HTML]{FEF9F9}1 & \multicolumn{1}{c|}{\cellcolor[HTML]{57BB8A}20} & \cellcolor[HTML]{FFFFFF}0 & \multicolumn{1}{c|}{\cellcolor[HTML]{5BBD8D}79} & \cellcolor[HTML]{FFFCFC}1 \\ \hline

\hline
\end{tabular}
\end{table*}

For UC\#1, there are no statistical differences between Abis and Tims between the treatment groups (Control vs. Plugin). All participants in both treatments had high success rates. Tims had 100\% completion rates in both treatments. Abis in the Plugin group performed better than the Control group (94.7\% vs. 83.3\%), but the difference is not statistically significant. This reflects that UC\#1 was a simple enough use case, with the majority of Abis able to overcome the inclusivity bugs (Bug \#1 to Bug \#5) to complete the task. Recall, inclusivity bugs need not be show stoppers, but they add an additional cognitive tax every time a user faces them. 

For UC\#2 and UC\#3 in the Control group, Abis performed worse than Tims by about 33\%, with the difference being statistically significant (p-value $< 0.05$). However, there is no difference when we compare the Abis and Tims in the Plugin group. Both Abis and Tims have a 100\% completion rate for UC\#2 and 95\% for UC\#3. This suggests that our redesign helped Abis overcome barriers to completing these tasks.
%In UC\#2 all participants in the Plugin groups (Abis and Tims) successfully completed the task and for UC\#3 Abis' completion rate was 94.7\% as compared to 95\% for Tims. 

%and \textit{Task \#3 - Request help to solve the PR} (61\% (Abi) x 94\% (Tim), \textit{p-value = 0.016}).

% On the other hand, in the plugin group, there were no statistically significant differences, and users with Tim and Abi facets achieved similar performance (i.e., Task \#2 - 100\% (Abi) x 100\% (Tim) and Task \#3 94\% (Abi) x 95\% (Tim)).

%We evaluate the effectiveness of users of both groups in completing the tasks using the \textit{Chi-Square test} to check the independent relation between two categorical variables~\cite{josephus2021predict}. Our results indicate that there was an initial effectiveness gap between users with Tim and Abi's cognitive styles in two tasks (Table~\ref{tab:difference}). Abi's were statistically significantly worse than Tims in \textit{Task \#1 - Submit pull request} (i.e., 81\% (Abi) vs. 100\% (Tim), \textit{p-value = 0.04311445}) and \textit{Task \#3 - Request help to solve the PR} (62\% (Abi) x 90\% (Tim), \textit{p-value = 0.04859509}). After implementing the proposed fixes in the plugin, we could not find and statistically significant difference, and users with Tim and Abi facets achieved similar performance (e.g., Task \#1 - 100\% (Abi) x 96\% (Tim) and Task \#3 92\% (Abi) x 96\% (Tim)).

\begin{table*}[!hbtp]\scriptsize
\centering
\vspace{-2.5mm}
\caption{Effectiveness of tasks completed and comparison among groups.}
\vspace{-2mm}

\label{tab:difference}
\begin{tabular}{c|cc|cc|cccc}
\hline
 & \multicolumn{2}{c|}{\textbf{Abi}} & \multicolumn{2}{c|}{\textbf{Tim}} & \multicolumn{4}{c}{\textbf{Differences}} \\ \cline{2-9} 
\multirow{-2}{*}{\textbf{UC}} & \multicolumn{1}{c|}{\textbf{Control}} & \textbf{Plugin} & \multicolumn{1}{c|}{\textbf{Control}} & \textbf{Plugin} &
\multicolumn{1}{c|}{{\color[HTML]{F56B00} \textbf{Abi}}-C x {\color[HTML]{3166FF} \textbf{Tim}}-C} &
\multicolumn{1}{c|}{{\color[HTML]{F56B00} \textbf{Abi}}-P x {\color[HTML]{3166FF} \textbf{Tim}}-P} &
\multicolumn{1}{c|}{{\color[HTML]{F56B00} \textbf{Abi}}-P x {\color[HTML]{F56B00} \textbf{Abi}}-C} & 
\multicolumn{1}{c}{{\color[HTML]{3166FF} \textbf{Tim}}-P x {\color[HTML]{3166FF} \textbf{Tim}}-C} \\ \hline \hline

%83,3%	94,73%	100%	100%

\textbf{\#1} & \multicolumn{1}{c|}{\cellcolor[HTML]{77C8A0}83.3\%} & \cellcolor[HTML]{57BB8A}94.7\% & \multicolumn{1}{c|}{\cellcolor[HTML]{57BB8A}100\%} & \cellcolor[HTML]{5EBE8F}100\% & 
\multicolumn{1}{c|}{
%{\color[HTML]{FE0000} \textbf{$\Downarrow$}} -19\% 
%{\color[HTML]{32CB00} \textbf{$\Uparrow$}} 
%*
-} & \multicolumn{1}{c|}{-} & \multicolumn{1}{c|}{-} & - \\ \hline

\textbf{\#2} & \multicolumn{1}{c|}{\cellcolor[HTML]{62C092}61.1\%} & \cellcolor[HTML]{57BB8A}100\% & \multicolumn{1}{c|}{\cellcolor[HTML]{60BF90}94.4\%} & \cellcolor[HTML]{57BB8A}100\% & 
\multicolumn{1}{c|}{
{\color[HTML]{FE0000} \textbf{$\Downarrow$}} -33.3\% 
%{\color[HTML]{32CB00} \textbf{$\Uparrow$}} 
*
}
& \multicolumn{1}{c|}{-} & \multicolumn{1}{c|}{-} & - \\ \hline

\textbf{\#3} & \multicolumn{1}{c|}{\cellcolor[HTML]{96D5B6}61.1\%} & \cellcolor[HTML]{64C093}94.7\% & \multicolumn{1}{c|}{\cellcolor[HTML]{68C296}94.4\%} & \cellcolor[HTML]{5EBE8F}95\% & \multicolumn{1}{c|}{
{\color[HTML]{FE0000} \textbf{$\Downarrow$}} -33.3\% 
%{\color[HTML]{32CB00} \textbf{$\Uparrow$}} 
*
} & \multicolumn{1}{c|}{-} & \multicolumn{1}{c|}{
{\color[HTML]{32CB00} \textbf{$\Uparrow$}} +36.6\% 
%{\color[HTML]{FE0000} \textbf{$\Downarrow$}} 
*
} & - \\ \hline

\textbf{\#4} & \multicolumn{1}{c|}{\cellcolor[HTML]{CBEADB}27.7\%} & \cellcolor[HTML]{70C59B}89.4\% & \multicolumn{1}{c|}{\cellcolor[HTML]{CDEBDC}33.3\%} & \cellcolor[HTML]{57BB8A}100\% & \multicolumn{1}{c|}{-} & \multicolumn{1}{c|}{-} & \multicolumn{1}{c|}{
{\color[HTML]{32CB00} \textbf{$\Uparrow$}} +61.7\% 
%{\color[HTML]{FE0000} \textbf{$\Downarrow$}} 
**
} & 
{\color[HTML]{32CB00} \textbf{$\Uparrow$}} +66.6\% 
%{\color[HTML]{FE0000} \textbf{$\Downarrow$}} 
**
\\ \hline
\multicolumn{9}{l}{(* p$\leq$0.05; ** p$\leq$0.01)}
\\ \hline \hline
\end{tabular}
\vspace{-3.5mm}

\end{table*}

All participants struggled to complete UC\#4 in the Control group; Abis' completion rate was 27.7\% as compared to Tims' 33.3\%. The redesign helped both Abis and Tims, with Abis' improvements at 61.7\% and Tims' at 66.6\%. The improvements in the Plugin group compared with the Control group were statistically significant (p-value $<$ 0.001). This result highlights that designing an interface to improve the experience of one underserved population can help make the software better for the larger population.
% In the control group, we can notice that participants with Tim and Abi's cognitive styles struggled to complete \textit{Task \#4 - Upload file}. The bug fixes implemented in the plugin statistically significantly helped both---Tim (100\% (plugin) vs. 33\% (control), \textit{p-value $<$ 0.001}%.012772e-05}
% ) and Abi (89\% (plugin) vs. 27\% (control), \textit{p-value = 0.0001332142}).

\begin{figure}[!t]
  \centering
  \includegraphics[width=8cm]{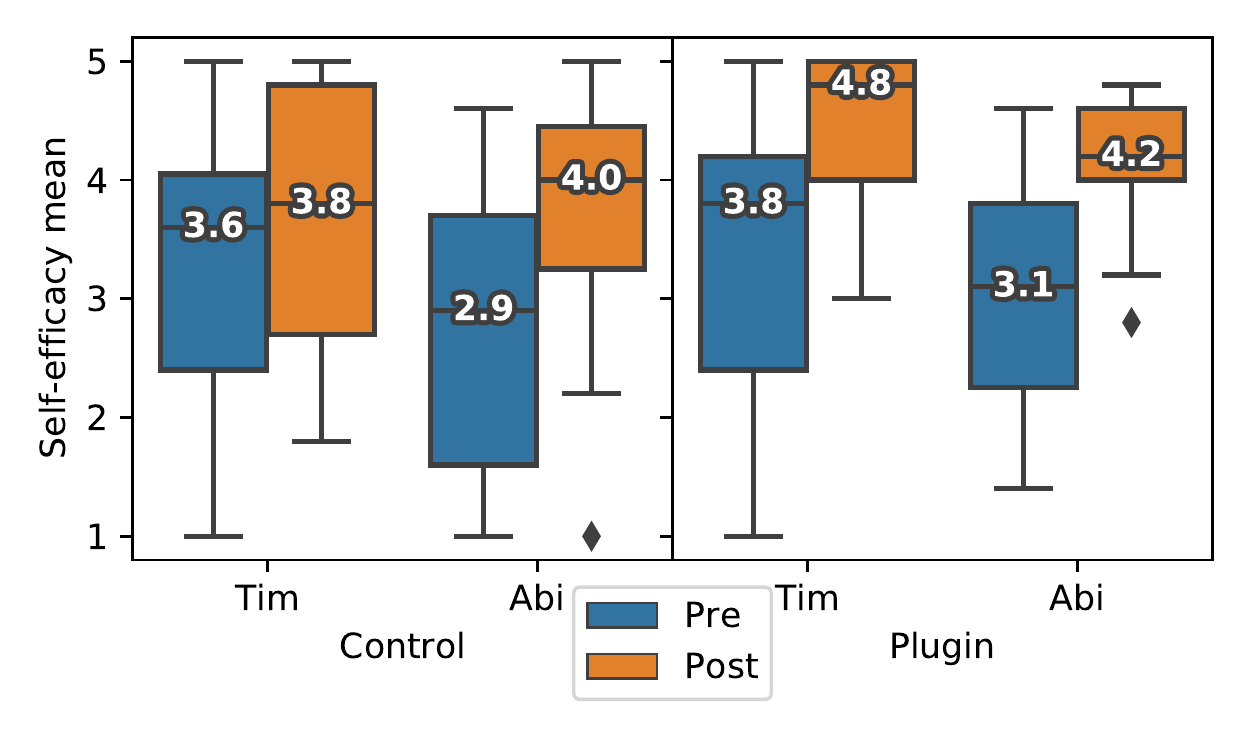}
  \vspace{-3.5mm}
  \caption{Self-efficacy results.}
 \vspace{-2.5mm}
  %\vspace{-7mm}
  \label{fig:selfefficacy}
\end{figure}

\textit{Impact on self-efficacy}. Figure~\ref{fig:selfefficacy} presents the results of the self-efficacy questionnaire that participants filled out at the beginning (`pre') and end (`post') of the study, disaggregated by treatment groups (`Control' and `Plugin') and per persona (`Tim' and `Abi'). 

At the beginning of the experiment (`pre'), Abis (2.9) had a lower self-efficacy as compared the Tims (3.6). After performing the experiment tasks (`post'), both types of participants gained confidence. Given these participants had never interacted with GitHub to submit a pull request before, it is expected that they were not confident in completing the tasks at the start of the experiment. But, after completing use cases (UC\#1 to UC\#3), their self-efficacy improved. It is heartening to note that the failure to complete UC\#4, the last use case, did not dampen those participants' starting self-efficacy.  
% Given the low self-efficacy values in the control group, our results suggest that the GitHub environment can be difficult for every newcomer, regardless of their GenderMag cognitive style (i.e., Abi or Tim). We can see that both control and plugin participants increased self-efficacy. 

The improvement in participants' self-efficacy was larger (`pre' vs. `post') in the Plugin group for both Abis and Tims. 
We calculated the Wilcoxon signed-rank test, a frequently used nonparametric test for paired data (e.g., pre- and post-treatment measurements)~\cite{rosner2006wilcoxon}, which indicates that the difference in improvement (`pre' vs. `post') between the Control and Plugin groups is significant for both types of participants; improvement for Abis has p-value $ = 0.005$, and Tims has p-value $< 0.001$. 
%is significant for the Tim plugin Pre vs. Tim plugin Post (\textit{p-value = 0.0009}) and for Abi plugin Pre vs. Abi plugin Post (\textit{p-value = 0.005}). 
We calculated Cliff's delta effect size measure~\cite{cliff1993dominance} to calculate the magnitude of these differences among the Plugin group. The effect size of improvement for Tims `pre' vs. `post' is large (delta = 0.682), as well as for Abis `pre' vs. `post' improvements (delta = 0.542). %0.6479592

%It highlights how the use of the plugin impacted an excellent experience for newcomers that are performing a pull request for their first time on the GitHub platform. Thus, it can contribute to making them continue using and start engaging more in OSS activities.

%\blue{

%\boldification{We discuss how the plugin interface helped participants overcome the challenges}

\textit{Impact of the proposed interface on participant experiences.} The questionnaire that participants filled out after every task (Section \ref{sec:assessing-inclusivity}) confirms that the control group participants faced more challenges. In the following, we discuss participants' reflections on what their difficulties when performing the experiment tasks and how the Plugin design helped them.

% \boldification{In UC\#1, Tim participants had difficulty finding the editor and README files}
In UC\#1, the main challenge Tims faced in the Control group was the difficulty of finding the editor and the README files (Bug \#1 of Table~\ref{tab:inclusivitybugs}). P44 mentioned ``\textit{Finding the README file, definitely, because I didn't know where to look for all these files, I didn't think it would be like in the middle of those files.}'' 
%
%\boldification{The interface solved this bug by presenting the README file information more explicitly}
The plugin solved this problem by presenting the README file information to users more explicitly (improving visibility) through a new tab called home. None of the participants in the Plugin group mentioned finding the README to be a problem. 

%\boldification{We also observed that the interface helped Abis through button colors and tooltips.}
The Abis in the Plugin group mentioned that the improved visibility of features in the redesigned interface (button colors, tooltips) helped them complete UC\#1. The tooltips allowed comprehensive information processors to gather the necessary information before starting the task. It also improved their self-efficacy by letting participants know they were on the right path. Indeed, P29 mentioned that ``\textit{the tooltip guides me into the execution of the task}''.
 
%\boldification{However, the plugin was not the definitive solution for Abis -- they still had difficulties associated with risk-averseness}
The redesigned interface, however, did not help Abi-like participants in figuring out how to edit the file and save it (Bug \#4). P1 said, ``\textit{Starting the Edit process was really hard. And once you have a little computer knowledge and you actually get into the Edit tab, you can look at the various files you want to edit and then go through the process}''. This comment highlights Abi's risk-averseness when having to use new features.

%\boldification{In UC\#2, the plugin helped both Abis and Tims to find the changed files}
In UC\#2, a difficulty that both Abi- and Tim-like participants faced in the Control group was finding the changed files in the pull request interface (Bug \#6). Tims and Abis in the Plugin group mentioned that the changed files in the navigation menu (improved feedback) and the tooltips (improved visibility) helped them to complete the task. This is an example of how a solution designed to help one class of users (Abis) helps a broader population (also Tims). 
%Tim users in the plugin group mentioned that the interface helped with clear labeling, the tooltip, the icons, and the navigation menu. 
%As in UC\#1, Tims in the Control group showed more satisfaction with regular GitHub interface elements.

%\boldification{In UC\#3, participants had difficulty to find how to request help in the Control group}

The main difficulty in UC\#3 was finding out how to request help. Some participants reported that their first idea was directly contacting the experienced user. P43 mentioned: ``\textit{I thought there would be a way that I could just like leave them a personal message and ask for help rather than posting. It [comment in the interface] looks like a public comment.}'' Other participants tried to contact the user directly by going to their GitHub profile page and looking for a direct message option, which GitHub does not offer. A majority did not realize they could use the `@' button in the panel to direct their comments to a specific contributor.

%\boldification{The plugin helped most participants by including a mention icon}
Participants in the Plugin group used the tooltip associated with the mention icon (`@') to figure out this feature. The improved visibility of the `@' icon helps process-oriented learners, who would be hesitant to tinker around the interface to find and use the '@' button. With this fix, an Abi participant mentioned that the task was intuitive (P40): ``\textit{Once I recognized that I needed to do this task as well, it was pretty intuitive.}''. 
In UC\#4, participants in the Control group faced more difficulty figuring out how to obtain push access: one Abi participant and ten Tims mentioned having that difficulty. Only three of these ten participants overcame this challenge and successfully completed the task. None of the participants mentioned this challenge in the Plugin group. Abis in the Plugin group mentioned that the improved visibility afforded by the green fork button and the feedback message was helpful. P26 said: ``\textit{interface messages when trying to upload the file helps a lot}''. Tims in the Plugin group said the same, exemplified by P5: ``\textit{So when I went back, I saw that the fork was highlighted in like the same green color. (...) It really just puts me back in the right direction}''.

\rqtwo[
    \tcblower
    \textbf{Answer:} A GenderMag-inspired redesign of the GitHub interface removed the task completion gaps between Abi and Tim participants for UC\#2 and UC\#3. For UC\#4, the redesign significantly improved task completion for both Abi and Tim participants.
    %We were able to remove the gap between users with Tim versus Abi cognitive styles in \textit{Task \#2} and \textit{Task \#3}. In the control group, Abi users performed worse than Tim users, while in the plugin group, people with both cognitive styles achieved similar performance. For \textit{Task \#4}, people with both cognitive styles struggled, and our changes helped both. Hence, fixing inclusivity bugs  can help all users. 
   % The qualitative data highlighted that participants from the control group faced more challenges than those in the plugin group. The interface elements included by the bug fixes helped them complete the tasks.
]{}

%Although it is not a perfect solution yet, our study can make users with different cognitive styles thrive more when they attempt to contribute to an OSS project. Our results also suggest that by attempting to remove gaps for a minority of users in a software platform (e.g., GitHub), we improved the overall performance of different users, highlighting the importance of the development of software considering the diverse aspects of the end-users. 

\section{Discussion}
\label{sec:discussion}

A decade of research on gender HCI has found that individual differences in how people problem solve---how they think when interacting with a software---cluster by gender. Past research has shown that current software and documentation embed inclusivity bugs---bugs that disproportionately affect a subset of users whose cognitive styles are unsupported by the software. These inclusivity bugs result in an additional cognitive tax every time a user faces the bug, which can add up to create barriers to participation.  

%\boldification{Our study found inclusivity bugs in GH}

In our study, we investigated to what extent the GitHub interface embedded inclusivity bugs and how these inclusivity bugs impacted users' performance with different cognitive styles (i.e., Abis and Tims). After applying the GenderMag method on GitHub, we found 12 inclusivity bugs that affect the Abi persona. 

%\boldification{We found similarities between the bugs we found and the bugs identified in previous work.}
\textit{Alignment with past research.}
Our findings are similar to that of past GenderMag research identifying inclusivity bugs in OSS projects.
Padala et al.~\cite{padala2020gender} found Information Processing, Self-efficacy, and Learning Style facets favored by Abi to be the most frequent facets that were not supported by OSS projects, and the lack of support of these facets was instrumental in causing the top reported barriers to contribute. More specifically, they found that:
(1) comprehensive information processors would feel disoriented because of insufficient upfront information provided in the project README. In our study, Abi-like participants also reported feeling lost in the Control group; 
(2) participants with lower computer self-efficacy
were worried about completing the task and described a lack of knowledge of the technologies as a reason for it. These findings also appear in our results---participants in the Control group felt scared by the GitHub interface; % during their attempt to start contributing to OSS projects. 
(3) process-oriented learners were hampered by a lack of clear instructions on how to contribute. We observed that Abi-like participants in the Control group also got stuck completing some of the tasks because of a lack of instructions on how to use many of the GitHub features. %\italo{The learning style cognitive facet spectrum ranges from process-oriented learning to learning by tinkering and defines how an individual approaches learning new technology~\cite{burnett2016gendermag}. When learning new technologies, process-oriented learners prefer to start with a process, like tutorials or other step-by-step communications/structures. Instead of learning a process, individuals who learn by tinkering (tinkerers) construct their understanding of the problem bottom-up by tinkering in the problem space~\cite{beckwith2006tinkering}
% .}

Fixing these inclusivity bugs not only helps Abi-like users, whose facets were used to redesign the software but can also make the software better for the larger population. Vorvoreanu et al.~\cite{vorvoreanu2019gender} in their work found that a redesign of their software to fix the inclusivity bugs found via GenderMag helped women do better (who had twice the failure rate as men in `pre-fix' version), removing the gender gap in the `post-fix' version. Moreover, both men and women participants had fewer failures. 
We found similar results, where redesigning the GitHub interface to accommodate Abi-like users also helped the Tim-like participants in our study (66.67\% improvement among Tims in UC\#4).
%Concerning participants with Tim's persona, removing the inclusivity bugs also helped them face fewer challenges in some situations. These results highlight that by attempting to make software less gender-biased, we help improve the performance of all participants with different cognitive styles.

%\boldification{These inclusivity bugs become gender bugs because of distribution of men/women to Tim/Abi}
\textit{Cognitive diversity bugs can become gender-bias bugs}.
Past research using GenderMag has shown that the inclusivity bugs created when Abis' cognitive styles are unsupported also become gender-bias bugs because individual differences in how people problem solve cluster by gender~\cite{beckwith2006tinkering, burnett2011gender, burnett2010gender, burnett2016gendermag}. In our data set, we see that the distribution of Tim facets aligned more closely with the distribution of men. We had 63\% men who had a majority of Tim facets compared to those who had a majority of Abi facets. In our study, perhaps due to the small sample size and our recruitment pool, we had an equal distribution of Abis and Tims among the women participants.

\textit{The need to make OSS tools and technology inclusive.}
OSS has a severe gender diversity imbalance, with the percentage of women ranging around 10\%. One of the challenges women face is a lack of sense of belonging, which may make them less inclined to share their opinions with the rest of the team. We noticed such reticence among our women participants in opining about their difficulties. In contrast, the men (the majority comprised of Tim) in the study felt more empowered to talk about the challenges they faced and suggest how they would improve the GitHub interface. One reason for this difference in behavior can be because women tend to have lower computer self-efficacy than men within their peer sets~\cite{padala2020gender}. This can affect their behavior with technology~\cite{wang2018competence, burnett2011gender, cazan2016computer, huffman2013using}, indicating that women feel less comfortable sharing their opinions and are inclined to think that it is their fault for not being able to use a certain technology. By making the OSS tools and technology more inclusive, we can break the barriers that Abi-like users, typically women, face when using the tools and technology, which can add to their feelings of not belonging~\cite{trinkenreich2023belong} and impostor syndrome~\cite{trinkenreich2021women}. %This can in turn help reduce the diversity imbalance OSS currently faces.

%\boldification{Removing GH inclusivity bugs is especially important to help diverse newcomers onboard OSS projects}
Making GitHub inclusive of varied cognitive skills is important for OSS to attract newcomers. Making GitHub inclusive will remove additional barriers newcomers face when their cognitive styles are not supported by the tool~\cite{padala2020gender}.
When the gap between newcomers' skills and those needed to accomplish the task is too broad, it demotivates newcomers, causing them to drop out~\cite{balali2020recommending, steinmacher2015understanding}. This can particularly impact students who are still developing their skills and have limited time and experience when first contributing to an OSS project. 

%It also highlights the importance of the GenderMag method to find inclusivity bugs that could help a diverse set of users of a software product overcome cognitive style barriers that make them feel insecure about how to complete a task in an OSS environment. 

\section{Implications}
\label{sec:implications}

%In this section, we discuss possible research directions that emerged from our results and implications.

\textbf{\textit{Implications for social coding platforms}}. For the designers and developers of GitHub and other social coding platforms, our results highlight the importance of developing software that encompasses the diversity of users. Social coding platforms can insert inclusivity biases that are crosscutting to a large number of projects. Social coding platform designers should consider newcomers' cognitive styles to understand how they process information or use the technology itself and how they can accomplish tasks to help them reach their main goals. A more inclusive design means including more users by making it easier for them to contribute to OSS projects.

\textbf{\textit{Implications for Maintainers of OSS projects}}. Our work reports inclusivity bugs newcomers can face and what part of a task they can get stuck on. Maintainers can use this information to consider how they could mitigate these challenges. One suggestion would be to provide more information in the README/Contributing.md files. We also hope our work can foster and ignite the interest in OSS communities to investigate and remove inclusivity bugs in the different tools and technology they use. %More in-depth research would be needed to propose other solutions to address those challenges and evaluate if other new ideas can help solve the inclusivity bugs. 

\textbf{\textit{Implications for newcomers (Abis and Tims)}}. Our results are important for newcomers. We showed the difficulties they face, where they struggle most, and how the interface can help them. 
Abis, who notoriously have low self-efficacy, should be aware that the interface was not designed for their cognitive style, and poor performance is a reflection of the tool failing them and not a reflection on their self-worth or capability. Tims should be aware that developers with diverse cognitive styles exist and respect the differences.  
%Our study highlights the need for inclusion during the development process to increase the diversity of users that can interact with it by fixing inclusivity bugs that can affect users' efficacy.

\textbf{\textit{Implications for educators}}. Familiarizing students with the OSS contribution process is becoming more common~\cite{pintoFSG17}. Contributing to a real project helps students gain  real-life experience and allows them to add this experience to their resume, which aids them in securing jobs. Our results highlight that based on their cognitive styles, some students can face more challenges when interacting with the GitHub platform. Educators should understand those challenges and teach students how to overcome them. They can also explore other ways to facilitate students' learning of the GitHub platform.

\section{Limitations}
\label{sec:threatstovalidity}

Our investigation also has threats to validity and limitations. We focused our analysis on finding inclusivity bugs for newcomers based on GenderMag Abi's persona. We followed the guidelines suggested by Hilderbrand et al.~\cite{hilderbrand2020engineering} and focused on this persona because its facet values tend to be more undersupported in software than the other personas~\cite{guizani2022debug, burnett2016finding}. However, fixing problems from only this persona's perspective could leave non-Abi newcomers less supported. This was a clear trade-off that could impact Tims, for example. However, the results from out experiment that include both Tim and Abi personas, showed that the performance of the Tim participants also improved for some tasks.
\balance

Despite our best efforts to recruit women for the experiment, there is a gender imbalance in the sample. At the same time that having more women would be important for the gender balance perspective, we would lose in terms of representativeness of the population of interest. Still, although the number of women is lower than men, we have almost the same amount of Abi (37) and Tim (38) participants. Nevertheless, this paper aims to investigate the cognitive facets, and some men also present facets associated with Abi's persona. 

In the GenderMag analysis, we carefully conducted the walkthroughs on GitHub following the procedures described by Burnett et al.~\cite{burnett2016gendermag}. We had different meetings to review the GenderMag analysis and solutions proposed to fix the inclusivity bugs and the members of our research group had previous experience in conducting GenderMag analysis. Another concern is that the GenderMag method only relies on participants' gender, though that is not the case. Vorvoreanu et al.~\cite{vorvoreanu2019gender} states that the keys to more inclusive software lie not in someone's gender but in the facet values themselves. As this answer makes clear, GenderMag can be used to find and fix inclusiveness issues without ever speaking of gender.

%To improve the experiment instruments (e.g., questionnaires and task definitions) before experimenting with the real participants, we conducted a pilot study with five researchers outside our group to collect feedback. We did an iterative process where we conducted a first pilot study session. Then we applied the changes needed and ran other pilot study sessions until we agreed that the instruments were reliable enough to start the actual study with our target population. 

%In our experiment, we needed to select participants to draw general results for a more significant population in OSS projects. Due to that, we decided to run the study initially as individual sessions. To include more participants, we opted to optimize the study by asking for professors to apply the study during their classes, where the participants had to follow the instructions provided and perform the tasks requested in the study. To mitigate any threat in this new environment, participants were in the same room and could ask questions to the professor if they needed any help or did not understand the tasks required. We collected additional 38 valid responses to include in the final analysis.

We recruited 75 undergraduate students from diverse STEM majors from 5 different universities in the US and Brazil. Most participants were pursuing Computer Science majors. We acknowledge that the sample is not representative of the population under analysis. But, we decided to not seek for generalization, but to understand the phenomenon in a controlled environment that would generate initial evidence to be further investigate. Therefore, future studies may investigate whether newcomers from different countries or with education levels to compare the results.

Regarding the plugin development and evaluation, we ran tests during the development to assert its usability and correctness. However, the plugin could have different behaviors depending on the browser. To mitigate this threat, we made available a pre-configured computer in case the plugin did not behave as we expected during the experiment.

We collected the time participants spent completing the tasks. However, the high number of participants that did not complete the tasks made it hard to compare the time differences between groups. Future studies with larger samples may help to investigate time differences.

Concerning the qualitative analysis, we are aware that data interpretation can lead to bias. To mitigate subjectivity, we employed two researchers who independently coded the answers and conducted meetings to discuss and resolve conflicts. 
Still, this qualitative piece was important to collect the feedback from the users during their activity. We chose to provide this more subjective understanding to complement and enrich our results, instead of collecting only objective data.

\section{Conclusion}
\label{sec:conclusion}

Making software products usable to people regardless of their differences has practical importance. If a project's development tools or products fail to achieve inclusiveness, not only does its adoption fall but so does the involvement of underrepresented populations in the teams themselves~\cite{ford2016paradise, mendez2018open}. In this work, we found 12 inclusivity bugs in the GitHub interface for four tasks that are common for OSS newcomers. These bugs mainly affect users with cognitive styles that are more common to women---defined in the Abi persona~\cite{burnett2016gendermag}. We proposed fixes to the inclusivity bugs, implemented them in a plugin that changed the GitHub interface, and evaluated them through a between-subject experiment with 75 newcomers. 

%\italo{Abi users have cognitive styles that are unsupported by the original design, with the control group being able to complete only 67\% of the tasks. While in the plugin, the percentage increased to 95\%, indicating that the redesign improves the easiness of use.} 

We found that Abi participants in the Control group (regular GitHub) underperformed Tim participants in some use cases, with Abis in the Control group completing only 67\% of the tasks. Implementing the fixes based on the GenderMag analysis reduced these differences and improved the performance of Abi participants to 95\%, indicating that the redesign improved GitHub's usability and learnability. In one of our use cases, both Tim and Abi participants faced challenges, and the bug fixes implemented in the plugin significantly helped both participants (66\% improvement). We also noticed an overall increase in the self-efficacy perception for both Abi- and Tim-like participants in the Plugin group, highlighting how solving inclusivity bugs for minorities can also help the majority population. 

In future work, %we intend to conduct other GenderMag analyses in different tasks on GitHub to investigate possible hidden inclusivity bugs and propose solutions to keep improving the interface and facilitate the onboarding of newcomers in OSS communities. Further, 
we plan to use our results to continue exploring the inclusivity barriers in tools and infrastructure to improve newcomers' performance and make tools and projects more friendly for those who want to engage in OSS projects.

\section*{Acknowledgment}

This work is partially supported by the National Science Foundation under grant numbers 1900903, 1901031, 2236198, 2235601, CNPq \#313067/2020-1, CNPq/MCTI/FNDCT \#408812/2021-4, and MCTIC/CGI/FAPESP \#2021/06662-1. We also thank the students for participating in our study and Zachary Spielberger for helping develop the plugin.

%1815486, 1815503, 
%This work is partially supported by the National Science Foundation under Grant numbers 1815486, 1815503, 1900903, and 1901031, CNPq grant \#313067/2020-1. We also thank the students for participating in our study and Zachary Spielberger for helping with the plugin development.

%This work is partially supported by CNPq/MCTI/FNDCT (grant \#408812/2021-4) and MCTIC/CGI/FAPESP (grant \#2021/06662-1).

%\section*{Acknowledgments}
%The National Science Foundation partially supports this work under Grant numbers 1815486, 1815503, 2008089, 1900903, and 1901031, CNPq grant \#313067/2020-1. We also thank the students for participating in our study and Zachary Spielberger for helping with the plugin development.

%\newpage

\bibliographystyle{IEEEtran}
\bibliography{IEEEabrv,bibtex.bib}

\end{document}